# Simulation study of a photo-injector for brightness improvement in Thomson scattering x-ray source via ballistic bunching

DING Yun-Ze[1,2;1)]    DU Ying-CHao[1,2;2)]    ZHANG ZHen[1,2]    HUANG Wen-Hui[1,2]

1 (Department of Engineering Physics, Tsinghua University, Beijing 100084, China)

2 (Key Laboratory of Particle & Radiation Imaging (Tsinghua University), Ministry of Education, Beijing 100084, China)

**Abstract:** Increasing peak brightness is beneficial to various applications of Thomson scattering x-ray source. Higher peak brightness of scattered x-ray pulse demands shorter scattering electron beam realized by beam compression in electron beam-line. In this article, we study the possibility to compress electron beam in a typical S-band normal conducting photo-injector via ballistic bunching, through just adding a short RF linac section right behind the RF gun, so as to improve peak brightness of scattered x-ray pulse. Numerical optimization by ASTRA demonstrations that peak current can increase from ~50 A to >300 A for a 500 pC, 10 ps FWHM electron pulse, while normalized transverse RMS emittance and RMS energy spread increases very little. Correspondingly, the peak brightness of Thomson scattering x-ray source is estimated to increase about 3 times.

## 1. Introduction

Thomson scattering x-ray source with high peak brightness, proposed and demonstrated experimentally in 1990s' [1,2], has been studied and developed worldwide. The Accelerator Laboratory of Tsinghua University has also designed and built such an x-ray source (TTX) [3-5]. Thomson scattering x-ray source generates high brightness, quasi-monochromatic, ultra-short x-ray pulse by scattering a laser beam off a relativistic electron beam, with applications in atomic, nuclear, particle physics, medical field and so on. The peak brightness $B_x$, an important figure of merit of x-ray source performance, is defined as the number of photons/second/unit area/unit solid angle/unit bandwidth. In the case of a $180\,°$ interaction, based on some assumptions that are not so harsh, $B_x$ (photos/s/mm$^2$/mrad$^2$/0.1%b.w.) can be expressed as [6]

$$B_x = 5.05 \times 10^{18} \gamma^2 \frac{\lambda Q_e W_\gamma}{\Delta t_e \varepsilon_{ns}^2 x_L^2}. \tag{1}$$

[1)] E-mail: dingyz10@mails.tsinghua.edu.cn

[2)] E-mail: dych@mail.tsinghua.edu.cn

where $\gamma$, $Q_e$, $\Delta t_e$ and $\varepsilon_{ns}$ are the Lorentz factor, total pulse charge (nC), RMS bunch duration (ps) and normalized RMS emittance (mm-mrad) of the electron beam; $W_\gamma$ and $x_L$ are the total pulse energy (J) and RMS spot size (μm) of the incident laser beam; $\lambda$ is the central wavelength of the scattering x-ray pulse (μm). From Eq.(1), it's easy to know that using high brightness electron beam taking short pulse duration and low emittance at the same time is beneficial to improve the peak brightness of Thomson scattering x-ray source.

As mentioned above, to produce brilliant x-ray pulse in Thomson scattering source, high brightness electron beam is necessary. However, in conditional photo-injector, electron beam with relatively large total charge (e.g. 100 pC to 1 nC) could not keep short pulse duration because of intense space charge force. Therefore, beam compression is required in the systems having demand for high brightness. Two common compression methods are magnetic compression and velocity bunching. In magnetic compression scheme, no bunching is taken in the injector and beams are totally compressed after injector by one or more chicanes. Because of the use of long chicane and need of another RF linac section to compensate energy spread induced for compression, the device is complex and electron beam-line will be much longer than original one. Besides, beam emittance would degrade seriously during bending in the chicane when magnetic compression is applied to the case of low energy. Velocity bunching (or termed phase space rotation, PSR) proposed by Serafini and Ferrario, is another option for compression [7, 8]. Electron beam is injected into the first long booster linear accelerator at the zero acceleration phase and slips back in phase up to the peak acceleration phase, it can be compressed strongly as far as the extraction happens near the synchronous velocity. PSR is effective in bunching and can avoid serious emittance degradation by surrounding the whole bunching area with proper solenoid field to realize emittance compensation. However, the energy gain in the first booster linac is much less than nominal case without compression, so that extra accelerator may be required to satisfy the demand for beam energy. It also may enlarge the dimension of original system.

Actually, velocity bunching could be carried out based on another configuration called ballistic bunching [9, 10]. Ballistic bunching can be viewed as the "thin lens" version of phase space rotation. A positive, nearly linear energy chirp is imparted to electron beam after it passes through a short, high accelerating gradient RF cavities section inside which the synchrotron motion is very limited, and the chirp generates compression during the beam propagation along a drift space after the RF cavities section. In photo-injector, there is a long drift section between RF gun and the first booster linac according to emittance compensation theory [11, 12]. It's possible to compress beams in this drift section using ballistic bunching. A normal conducting S-band photo-injector is an often choice for Thomson scattering x-ray source. As an example, in this article we study the possibility to ballistic compress beam in this kind of photo-injector through merely adding a short RF linac section immediately after the electron gun to the original beam-line. In section 2, the beam-line configuration is depicted at first, and the mechanism of ballistic bunching and emittance compensation in photo-injector are reviewed then. In section 3, the feasibility of the configuration stated in section 2 is discussed with ASTR [13] simulation. Optimization result shows that to a 500 pC, 10 ps FWHM electron beam, the RMS beam length can be shorten over 3.5 times (pulse peak current increases from 50 A to >300 A) with very little normalized transverse RMS emittance and energy spread increasing. Correspondingly, the peak brightness of Thomson scattering x-ray source is estimated to increase about 3 times. Under this design, even to a system in commission, beams could be bunched primarily in the photo-injector without any changing to the original system except adding a short element. If the demand to pulse duration are not so short, this compression scheme may be a more convenient choice than magnetic bunching and PSR, avoiding complex design to a new magnetic compression system and enlargement of the system dimension.

## 2. Beam-line configuration

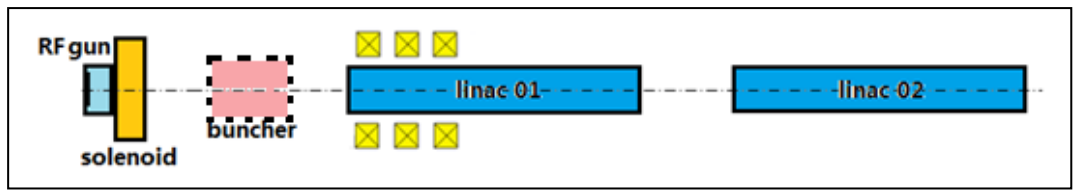

Fig. 1. (color online) The lay-out of the photo-injector considered in this article.

Injector based on a normal conducting S-band photocathode RF gun, one of the most practical designs for high brightness beam, is described schematically in Fig.1. An S-band BNL/KEK/SHI style 1.6 cell RF gun with a Cu photocathode inside serves as the electron source, followed by two 3-m S-band SLAC-style travelling wave sections to boost the beam to the required energy. A solenoid placed right behind the RF gun and solenoids surrounding the front half of the first linac section control the transverse size of beam and satisfy the requirement of emittance compensation.

According to theoretical description of the emittance compensation process, in order to damp emittance oscillations to get low emittance beam as it's accelerated to a high energy, beam need to be injected into the booster linac at an envelope waist (transverse RMS divergence $\sigma'$=0) and the envelope have to be matched to the accelerating gradient of linac and the focusing strength so as to stay close to a so-called invariant envelope (IE) given by [12]

$$\sigma_{IE} = \frac{1}{\gamma'} \sqrt{\frac{2I_p}{I_A \left(1 + 4\Omega^2\right)\gamma}} \, . \tag{2}$$

where $\gamma'$ is the normalized accelerating gradient defined as $\gamma'=eE_{acc}/m_ec^2$, $E_{acc}$ is the accelerating field, $I_p$ is the beam peak current in the bunch, $I_A$=17 kA is the Alfven current, and $\Omega^2$ equals to $eB_{sol}/mc\gamma'$ to a traveling wave RF structure. In the photo-injector described above, beam leaving the RF gun is focused immediately by the solenoid after the gun and passing over a drift space its envelope reaches to a waist and match $\sigma_{IE}$ at the entrance of the first booster linac.

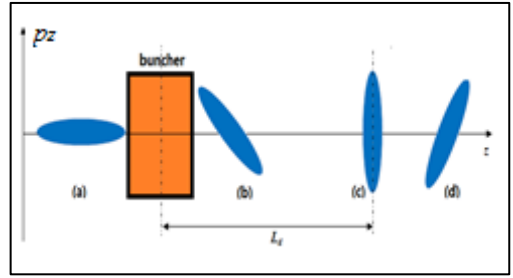

Fig. 2. (color online) The evolution of beam longitudinal phase space distribution ($z$-$p_z$) in the process of ballistic bunching. (a) is for the distribution before injection into the buncher; (b) is for the distribution just at the exit of the buncher; (c) is for the maximum compression; (d) is for the over-compression.

The unique difference in our lay-out from the common S-band photo-injector is a short linac section placed behind the gun solenoid (showed in Fig.1 by dashed lines) as a RF buncher to provide velocity modulation for the beam out of the gun relying on which to ballistic bunch the beam in the drift section after the buncher. The process of ballistic bunching can be illustrated with Fig.2. Beam is injected into the high gradient, short linear buncher near the zero acceleration phase. After leaving the buncher, a negative longitudinal monument tilt (or termed a positive chirp) is imposed over the length of the beam like Fig.2 (b) showing. Considering the interaction of an electron with the sinusoidal accelerating RF wave $E_z$=-$E_b$sin$\phi$ ($E_b$ is the peak accelerating field of the wave) and neglecting phase advance inside buncher, the chirp after the buncher can be described as [9]

$$\frac{d(\delta\gamma/\gamma)}{dz} = \frac{d(\delta\gamma/\gamma)}{dz}\bigg|_0 - k\frac{\Delta E_m}{E_0}\left(\frac{\cos\phi}{1+\frac{\Delta E_m}{E_0}\sin\phi}\right). \qquad (3)$$

where $E_0$ is the beam energy before entering the buncher, $\Delta E_m$ is the maximum gain available in the buncher, $k$ is the RF wave number, and $\phi=kz-\omega t+\phi_0$ is the RF phase. As a result, in the drift after buncher, particles at tail of the beam "catch up" particles at head gradually. When the pattern of the beam in longitude phase space becomes as Fig.2(c) showing, the beam achieves the maximum compression. As space charge force is neglected, the drift length needed for the strongest bunching could be estimated with the chirp from Eq.(3) by [9]

$$L_d = \frac{\beta^2 \gamma^2}{\left| \frac{d(\delta\gamma/\gamma)}{dz} \right|} . \tag{4}$$

If not promptly injected into a booster to freeze down de-bunching due to residual longitude space charge force, the beam will be "over-compressed" as depicted in Fig.2 (d). As stated above, ballistic bunching can be treated as a thin lens version of phase space rotation method. Different from the long RF structure version of velocity bunching, the beam is extracted still close to the zero acceleration phase and compressed mainly in the drift section after the buncher. Taken as an example, considering a 5 MeV beam with *3ps* pulse duration and 25 keV energy spread going through a 30 cm RF buncher whose peak longitudinal accelerating field is 35 MV/m at the zero acceleration phase, the drift from the buncher to longitudinal focus is about 0.97 m according Eq.(3) and Eq.(4). In fact, it will be a little longer than this because of the defocusing of longitudinal space charge force. Moreover, if the beam is not bunched so intensely that it can keep quasi-equilibrium without obvious crossover in longitude phase space, emittance growth could still be suppressed through adjusting field strengths of the gun and the linac solenoid to match the beam envelope to IE. In the following sections, the feasibility of this compression scheme will be discussed with the ASTRA simulation results.

## 3. Numerical simulation by ASTRA

Some numerical simulations about producing relatively high charge, ultra-short pulses with the lay-out stated in section 2 were carried out by code ASTRA. In this section, the simulation results are presented and the feasibility of the compression scheme is discussed.

### 3.1 Nominal case

Tab.1 Detailed beam-line parameters without buncher.

| element | parameter | value |
|---------|-----------|-------|
| RF gun | maximum gradient/MV $\mathrm{m}^{-1}$ | 100 |
| | driven laser radius/mm | 0.9 |
| gun solenoid | length/cm | 22.5 |
| | center location/m | 0.214 |
| linac 01 | entrance location/m | 1.5 |
| | accelerating gradient/MV $\mathrm{m}^{-1}$ | 20 |
| | solenoids length/cm | 120 |
| linac 02 | entrance location/m | 5.5 |
| | accelerating gradient/MV $\mathrm{m}^{-1}$ | 20 |

In simulation, we consider a 500 pC total charge pulse produced. The maximum gradient at the surface of the gun cathode is set to 100 MV/m, which is a tradeoff between keeping system stability and reducing space charge force; thermal emittance is included of 0.9 mm-mrad per mm RMS of the driving laser pulse assumed a flat-top distribution with Gaussian edge in longitude direction and a uniform distribution in transverse direction. Tab.1 shows the detailed beam-line parameters. When the buncher doesn't work, i.e. no bunching happens in the injector, the optimized result to a 10 ps FWHM beam shows ~0.60 mm-mrad normalized transverse RMS emittance and ~50 A peak current (RMS pulse duration ~2.8 ps) at the exit of the second TW section.

### 3.2 Longitude compression

The buncher chosen in this article is a conventional traveling wave structure consisting of 9 cells and operating with $2\pi/3$ mode at 2.856 MHz (about 30 cm in length). The buncher is better to be placed as close as possible to the RF gun in order to minimize the extent of space-charge de-bunching. But in consideration of the existence of the gun solenoid and some beam diagnostic elements behind it, the buncher is located at $z$=0.6 m in the simulation. To all the simulation results displayed later, all launching parameters of the electron beam and the locations of other elements except the buncher are the same as the nominal case.

As buncher's structure and location (i.e. the length of the drift section between the buncher and the first booster linac) are specified, the final pulse duration or peak current is mainly decided by the accelerating field gradient and beam injection phase of the buncher. Choosing different buncher gradients and beam injection phases, different energy chirps shown in Eq.(3) are obtained after beam transiting the buncher. Correspondingly, different length of drift is needed to achieve the maximum compression. As the location of the first booster linac is fixed, the beam is bunched to different compression factor at the entrance of the first linac; and the bunching is frozen down rapidly once the beam is injected into the booster linac as a result of the fast gain in energy (Fig.3).

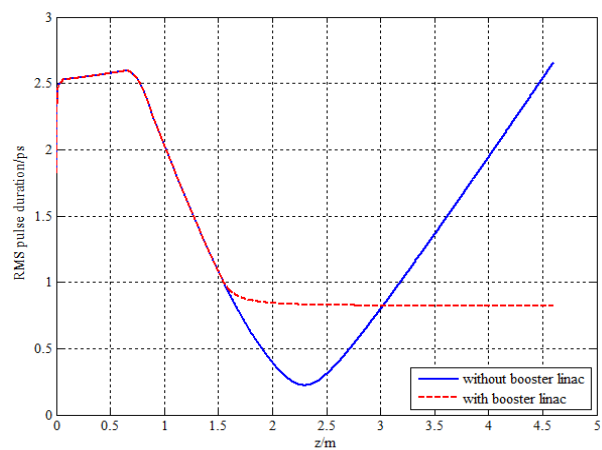

Fig. 3. (color online) Evolution of RMS pulse duration through the injector with the booster linac placed at 1.5 m (solid line), compared with over-compression when no booster existing (dashed line).

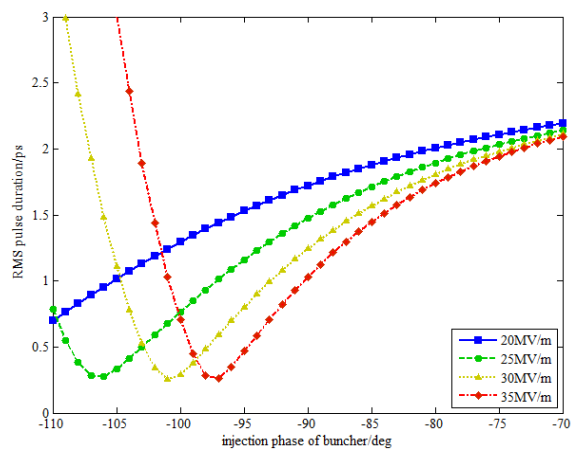

Fig. 4. (color online) RMS pulse duration at the exit of the injector as a function of the buncher injection phase with buncher gradient of 20 MV/m (blue, square), 25 MV/m (green, circle), 30 MV/m (yellow, triangle) and 35 MV/m (red, diamond).

The curves explaining the relationship between the longitude RMS length of the beam at the exit of the injector and injection phase of the buncher with different buncher gradients adopted are plotted in Fig.4. It's easy to understand that the higher gradient the buncher driven at, the stronger bunching force exerts on the beam so that the shorter pulse duration can be obtained at the same injection phase. Obviously we can't accept that the gradient of buncher needed by compression is too high, considering the saving of RF power and difficulty in power shading. In fact, injecting a beam into the buncher at more "negative" phases will also exert a stronger bunching force on the beam because of a slight deceleration of the beam. For instance, injecting beam at -106° into the buncher with 25 MV/m gradient gets an equivalent effect on longitudinal compression to injecting beam at -98° with 35 MV/m. From Fig.4, a wide range of beam lengths could be achieved through changing buncher's phase with different buncher gradients and injection phases adopted, while the shortest RMS pulse duration is < 300 ps. It's also worth noticing that RF gun work point (gun gradient and launching phase) also influences bunching process to some extent because of difference in energy spread beam possessing before injected into the buncher. Although optimization of the RF gun phase isn't discussed here, it has been proved that the scheme presented here is effective anyway. Moreover, in many cases, the beam duration got above is short enough.

### 3.3 Emittance compensation

It has been proven before that beam length can be compressed effectively with above scheme. But the demand for high brightness of electron beam requires beam not only to have high peak current but also to keep low transverse emittance. Optimization of transverse emittance can be accomplished by a careful tuning of gun solenoid field and the first linac solenoid field. To all simulation results presented in this section, beams launched phase is still chosen as same as the nominal case. Fig.5 presents the optimization result of the normalized RMS horizontal emittance for different longitudinal compression factors when the buncher is driven at 25 MV/m, 30 MV/m and 35 MV/m separately. As shown in Fig.5, when pulse compression factor is less than 4 (RMS beam length is not shorter than 700 fs), no matter what buncher gradient is chosen, the transverse emittance could be compensated very well. For example, evolution of the RMS pulse duration and normalized RMS horizontal emittance for a 0.8 ps compressed beam along the injector is presented in Fig.6, compared with those in nominal case. According to the emittance compensation theory, if beams can keep quasi-equilibrium without obvious crossover in longitude phase space before injected into the booster linac, emittance growth can still be well compensated through matching to IE by adjusting gun solenoid field and the first linac solenoid field. Fig.7 shows the longitudinal profiles of two beams with different compression factors at the entrance of the first linac. The profiles at the exit of the buncher are classified along the beam longitudinal length, and each colored point at the entrance of the first linac indicates the location in which it is at the exit of the buncher. It's obvious from Fig.7 that there is almost no overlap between different longitudinal slices of the beam compressed not so intensely, meaning the beam does not undergo crossover in longitudinal phase space so that the transverse emittance is able to be compensated well, while the growth of the emittance is notable because of the serious overlapping among several longitude slices at the head of the beam compressed more intensely.

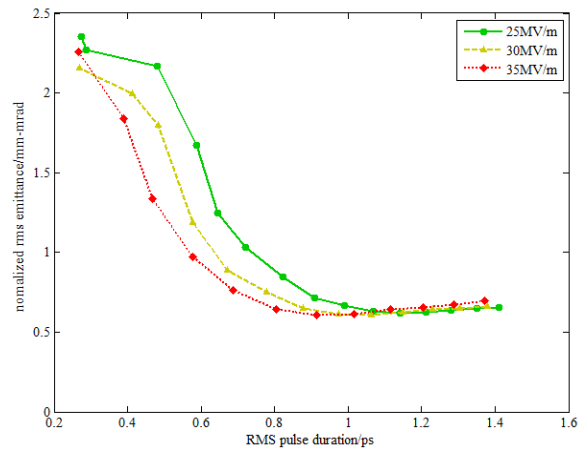

Fig. 5. (color online) Optimized normalized RMS emittance to different RMS pulse duration when buncher gradient is chosen as 25 MV/m (green, circle), 30 MV/m (yellow, triangle) and 35 MV/m (red, diamond).

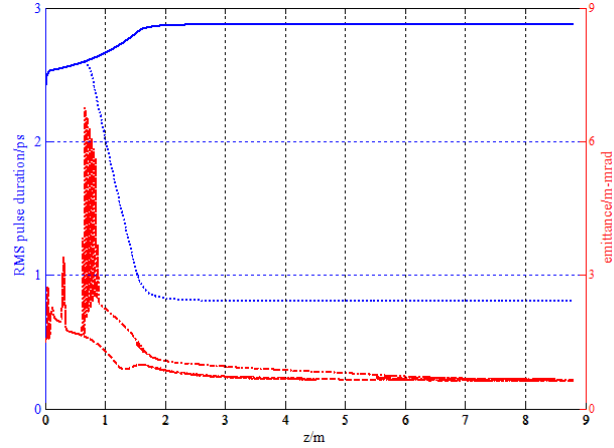

Fig. 6. (color online) RMS pulse duration (blue, left scale) and normalized RMS horizontal emittance (red, right scale) through the injector without (solid line and dashed line) / with (dotted line and dashed-dotted line) longitudinal compression.

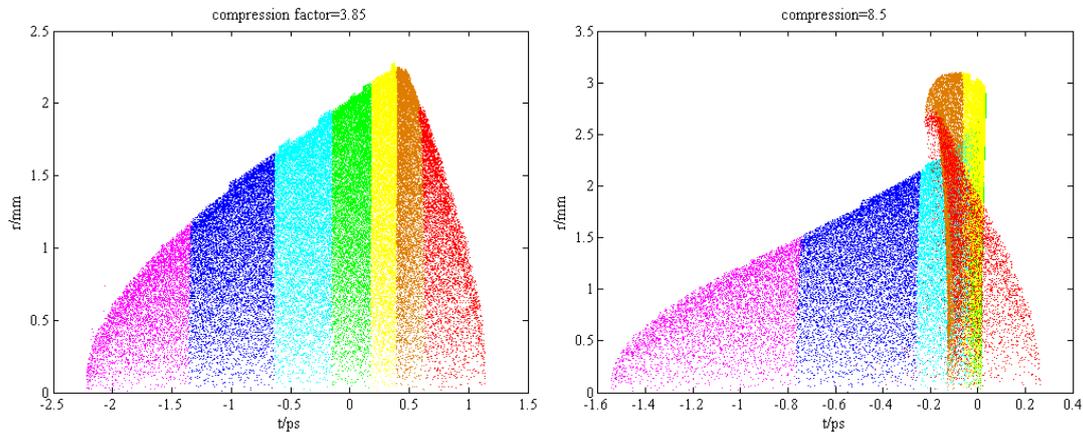

Fig. 7. (color online) Colored beam profile at entrance of the first linac with (left) / without (right) serious overlap between different longitudinal slices.

To several situations where emittance compensated well, detailed beam qualities at the exit of the injector are shown in Tab.2 ($E_b$=35 MV/m). The transverse slice emittance and slice current distributions are shown in Fig.8. As shown in Tab.2, relative high brightness beams with >300 A peak current and low transverse emittance and low energy spread could be achieved using the compression scheme presented. Furthermore, peak brightness of scattered x-ray pulse $B_x$ in each case is also estimated according to Eq.(1) and normalized with nominal case, shown in the last column of Tab.2. The peak brightness of scattered x-ray pulse becomes higher as the longitudinal focusing force rises. However, when compression factor arrives at a particular value, the peak brightness starts to decrease with a stronger focusing force, because of an abrupt degradation in transverse emittance. As a tradeoff between the emittance degradation and beam compressing, this scheme allows for a peak brightness of scattered x-ray pulse almost 3 times larger than that obtained in absence of ballistic bunching. In addition, the peak brightness calculated from Eq.(1) neglecting the influence of electron beam energy spread, which actually also impairs the peak brightness of scattered x-ray pulse. The relative energy spread shown in Tab.2 is gotten when the phase of the first and the second linac set to on-crest phase, which still keep low as nominal case. And the relative energy spread can even be decreased less by tuning the phases of both booster linacs.

Tab.2　Electron beam quality at the exit of linac 02 and corresponding normalized peak brightness of scatted x-ray pulse under different compression forces.

| compression factor | $E_k$/MeV | $\Delta z_{RMS}$/ps | $\varepsilon_{x,n,RMS}$/mm-mrad | $I_{peak}$/A | $\Delta E_{RMS}/E$ | $B_{x,n}$ |
|---|---|---|---|---|---|---|
| 0 | 125.5 | 2.80 | 0.60 | 56 | 0.12% | 1 |
| 3.15 | 125.3 | 0.91 | 0.61 | 238 | 0.17% | 2.96 |
| 3.58 | 125.1 | 0.81 | 0.64 | 304 | 0.15% | 3.00 |
| 4.19 | 125.0 | 0.69 | 0.76 | 412 | 0.14% | 2.49 |

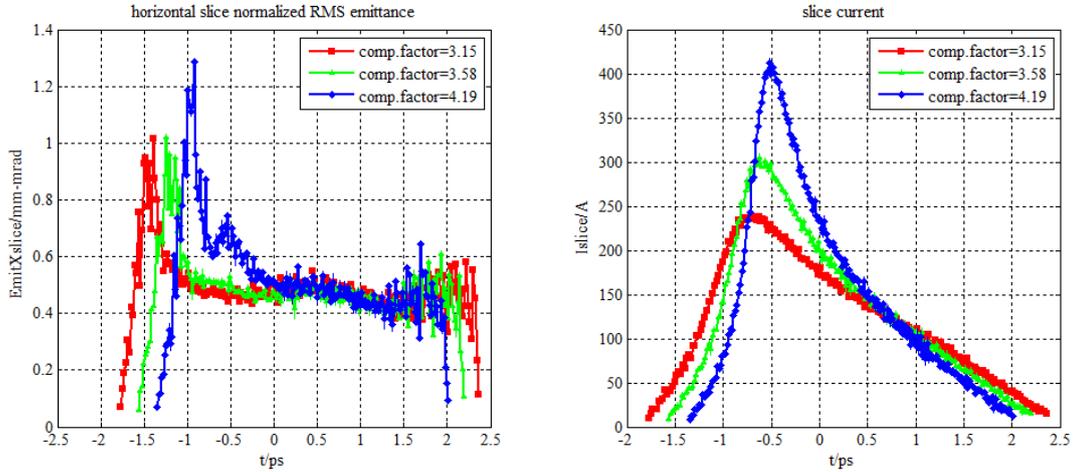

Fig. 8. (color online) Normalized RMS slice emittance (left) and slice current (right) for beams with different compression factors.

### 3.4 Jitter

To investigate the stability of the injector lay-out with ballistic bunching to operate as an electron beam injector for Thomson scattering source, the effect of RF amplitude and phase jitter has to be examined. It's possible that photocathode gun shares the same klystron with short buncher. The situation that errors of gun and buncher are coupled is also taken into account. Jitter simulation is based on the work point displayed at the third line in Tab.2. 300 simulations with 0.5 ° (RMS) RF phase random error and 0.1% (RMS) RF amplitude error were carried out for each situation. The RMS fluctuations of beam arrival time and some beam parameters at the exit of the injector are listed in Tab.3, compared with fluctuations in nominal case. It should be noticed that coupling of gun and buncher makes the beam parameters jitter much better than the option where the power is provided individually, but for arrival time jitter, this improvement is very limited. From the simulation result, under the RF stability level applied here, the fluctuations of longitudinal length and transverse emittance are obviously larger than those in nominal case. In sum, ballistic bunching scheme raise higher request on RF stability than nominal operation scheme, and powering RF gun and buncher with a same klystron performance better.

Tab.3    RMS variations of beam arrival time and some beam parameters at the exit of the injector.

|  | $t_{arrive}$ | $\Delta z_{RMS}$/fs | $E_k$/keV | $\Delta E_{RMS}$/keV | $\varepsilon_{x,n,RMS}$/mm-mrad |
|---|---|---|---|---|---|
| no bunch | 0.21° | 17.20 | 87 | 33.79 | 0.002 |
| uncoupled | 0.43° | 59.57 | 130 | 23.92 | 0.055 |
| coupled | 0.44° | 39.84 | 96 | 23.17 | 0.030 |

## 4. Conclusion

In this article, an idea to compressing electron beam utilizing ballistic bunching in the photo-injector for Thomson scattering x-ray source has been proposed. Just adding a short RF buncher right behind RF gun, bunching happens in the drift section between the gun and the first booster linac which originally exists in the photo-injector, so that the peak current increase effectively without enlarging the scale of the system. Numerical simulation by ASTRA studied the feasibility of this scheme. The result shows that transverse emittance does not appear obvious degradation when compression is not very intense, and the peak brightness of scattered x-ray is estimated to increase 3 times, as a tradeoff between the emittance degradation and beam compressing. Beam arrival time and beam parameters fluctuations from RF jitter can performance better when photocathode gun and RF buncher powered with the same klystron. Consequently, the ballistic bunching scheme proposed above has a great potential of using on photo-injector for Thomson scattering X-ray source to increase electron beams brightness. To systems demanding for several hundred amperes peak current, this scheme is a compact choice rather than magnetic bunching using chicane and phase space rotation using the whole length of the first booster linac to bunch, as while it's much easier to control transverse emittance and energy spread degradation. In addition, to photo-injector for other applications which requiring high peak current and compactness of system simultaneously such as THz radiation driven by high brightness electron beam, it's also an alternative scheme.